\documentclass[prd,superscriptaddress,twocolumn]{revtex4}

\usepackage{mathrsfs,amsmath}
\usepackage{eurosym}
\usepackage{amssymb}
\usepackage{bbold}

\usepackage{latexsym,epsfig,epstopdf}
\usepackage{amssymb,amsmath,amsthm}

\usepackage{times}
\usepackage{color}

\usepackage{lmodern}
\usepackage[utf8]{inputenc}

\newcommand{\ket}[1]{\left | #1 \right\rangle}

\newcommand{\braket}[2]{\left\langle #1|#2\right\rangle}

\usepackage{graphicx}
\usepackage{color}
\usepackage{braket}
\usepackage[mathscr]{euscript} 
\usepackage{perpage} 
\usepackage{slashed}

\usepackage{enumitem}

\def\be{\begin{equation}}
\def\ee{\end{equation}}
\def\bea{\begin{eqnarray}}
\def\eea{\end{eqnarray}}

\usepackage{calc}
\newsavebox\CBox
\newcommand\hcancel[2][0.5pt]{%
  \ifmmode\sbox\CBox{$#2$}\else\sbox\CBox{#2}\fi%
  \makebox[0pt][l]{\usebox\CBox}%  
  \rule[0.5\ht\CBox-#1/2]{\wd\CBox}{#1}}

\begin{document}

\title{Generalised Uncertainty Relations and the Problem of Dark Energy} 

\author{Matthew J. Lake}
\email{matthewjlake@narit.or.th}
\affiliation{National Astronomical Research Institute of Thailand, \\ 260 Moo 4, T. Donkaew,  A. Maerim, Chiang Mai 50180, Thailand}
\affiliation{Research Center for Quantum Technology, \\ Faculty of Science, Chiang Mai University, Chiang Mai, 50200, Thailand}
\affiliation{School of Physics, Sun Yat-Sen University, Guangzhou 510275, China}
\affiliation{Department of Physics, Babe\c s-Bolyai University, \\ Mihail Kog\u alniceanu Street 1, 400084 Cluj-Napoca, Transylvania, Romania}

%\affiliation{Frankfurt Institute for Advanced Studies, \\ Ruth-Moufang-Str. 1, 60438 Frankfurt am Main, Germany}, 
%\email{lake@fias.uni-frankfurt.de}

%%%%%%%%%%%%%%%%%%%%%%%%%%%%%%%%%%%
%%%%%%%%%%%%%%%%%%%%%%%%%%%%%%%%%%%
\begin{abstract}

We outline a new model in which generalised uncertainty relations, that govern the behaviour of microscopic world, and dark energy, that determines the large-scale evolution of the Universe, are intrinsically linked via the quantum properties of space-time. 
In this approach the background is treated as a genuinely quantum object, with an associated state vector, and additional fluctuations of the geometry naturally give rise to the extended generalised uncertainty principle (EGUP). 
An effective dark energy density then emerges from the field that minimises the modified uncertainty relations. 
These results are obtained via modifications of the canonical quantum operators, but without modifications of the canonical Heisenberg algebra, allowing many well known problems associated with existing GUP models to be circumvented. 

\end{abstract}


\maketitle

%%%%%%%%%%%%%%%%%%%%%%%%%%%%%%%%%%%
%%%%%%%%%%%%%%%%%%%%%%%%%%%%%%%%%%%
\section{Introduction} \label{Sec.1}

Thought experiments in phenomenological quantum gravity suggest that the canonical Heisenberg uncertainty principle (HUP) should be modified to incorporate both gravitational attraction and dark energy repulsion between fundamental particles \cite{Adler:1999bu,Maggiore:1993rv,Scardigli:1999jh,Bambi:2007ty,Bolen:2004sq,Park:2007az}. 
Heuristically, the HUP can be motivated by the Heisenberg microscope thought experiment \cite{Rae} and written as
\begin{eqnarray} \label{HUP-1}
\Delta x^i \, \Delta p_j \gtrsim \frac{\hbar}{2} \delta^{i}{}_{j} \, ,
\end{eqnarray}  
where $\Delta x^i$ and $\Delta p_j$ are the rather vaguely defined `uncertainties' in position and momentum. 
Extending this argument to include the effects of gravitational attraction between the observed particle and the probing photon leads to the generalised uncertainty principle (GUP),
\begin{eqnarray} \label{GUP-1}
\Delta x^i \gtrsim \frac{\hbar}{2\Delta p_j} \delta^{i}{}_{j} \left[1 + \alpha_0 \frac{2G}{\hbar c^3}(\Delta p_j)^2\right] \, ,
\end{eqnarray} 
where $\alpha_0 \sim \mathcal{O}(1)$ is a numerical factor of order unity \cite{Adler:1999bu,Maggiore:1993rv,Scardigli:1999jh}, whereas incorporating the effects of dark energy in the form of a cosmological constant $\Lambda \simeq 10^{-56}$ ${\rm cm^{-2}}$ \cite{Martin:2012bt} leads to the extended uncertainty principle (EUP),
\begin{eqnarray} \label{EUP-1}
\Delta p_j \gtrsim \frac{\hbar}{2\Delta x^i} \delta^{i}{}_{j} \left[1 + 2\eta_0 \Lambda (\Delta x^i)^2\right] \, , 
\end{eqnarray} 
where $\eta_0 \sim \mathcal{O}(1)$ \cite{Bolen:2004sq,Park:2007az,Bambi:2007ty}. 
Incorporating both effects implies the EGUP \cite{Kempf:1994qp,Kempf:1996ss}, 
\begin{eqnarray} \label{EGUP-1}
\Delta x^i\Delta p_j \gtrsim \frac{\hbar}{2} \delta^{i}{}_{j} \left[1 + \alpha_0 \frac{2G}{\hbar c^3}(\Delta p_j)^2 + 2\eta_0\Lambda (\Delta x^i)^2\right] \, . 
\end{eqnarray} 

In canonical quantum mechanics the HUP can be formulated more precisely using the Schr{\" o}dinger-Robertson relation \cite{Ish95} and the canonical position-momentum commutator, 
\begin{eqnarray} \label{HeisenbergAlgebra-1A}
[\hat{x}^i,\hat{p}_j] = i\hbar\delta^{i}{}_{j} \ \hat{\mathbb{I}} \, ,
\end{eqnarray}
which, together with the relations
\begin{eqnarray} \label{HeisenbergAlgebra-1B}
[\hat{x}^i,\hat{x}^j] = 0 \, , \quad [\hat{p}_i,\hat{p}_j] = 0 \, , 
\end{eqnarray}
form the canonical Heisenberg algebra, giving
\begin{eqnarray} \label{HUP-2}
\Delta_{\psi} x^i \, \Delta_{\psi} p_j  \geq \frac{\hbar}{2} \delta^{i}{}_{j} \, .
\end{eqnarray}
The inequality in Eq. (\ref{HUP-2}) is precise and $\Delta_{\psi} x^i$, $\Delta_{\psi} p_j$ are well defined as the standard deviations of operators acting on the probability density $|\psi|^2$. 
The subscript $\psi$ is used to emphasise this point and to distinguish these quantities from their heuristically motivated counterparts, $\Delta x^i$ and $\Delta p_j$. 

These considerations motivate the search for a modified form of quantum mechanics, which incorporates quantum gravitational effects, and which gives rise to a more precise formulation of the EGUP with well defined uncertainties derived from a well defined probability distribution. 
In the majority of the existing literature such a relation is obtained by introducing modified commutation relations, for example, of the form \cite{Tawfik:2014zca,Tawfik:2015rva}
\begin{eqnarray} \label{ModifiedCommutator}
[\hat{x}^i,\hat{p}_j] = i\hbar \delta^{i}{}_{j} (1 + \alpha\hat{\bold{p}}^2 + \eta\hat{\bold{x}}^2) \, \hat{\mathbb{I}} \, , 
\end{eqnarray}
where $\alpha$ and $\eta$ are appropriate dimensionful constants. 
Unfortunately, this approach leads to various pathologies including violation of the equivalence principle, violation of Lorentz invariance in the relativistic limit, the reference frame dependence of the `minimum' length, and the the inability to construct sensible multi-particle states, known as the soccer ball problem \cite{Lake:2018zeg,LakeUkraine2019,Lake:2019nmn,Lake:2020rwc,Lake:2021beh,Lake:2020chb}. 
These problems are well known \cite{Tawfik:2014zca,Tawfik:2015rva,Hossenfelder:2012jw,Hossenfelder:2014ifa}, but have remained essentially unsolved for over 25 years, ever since rigorous GUP models based on the modified commutator method were first formulated in the mid-1990's \cite{Kempf:1994su}. 

Recently, alternative GUR models have been proposed in which the canonical position and momentum operators are modified, giving rise to modified uncertainty relations including the EGUP (\ref{EGUP-1}), but in which the canonical Heisenberg algebra remains unchanged \cite{Bishop:2019yft,Bishop:2020cep}, except possibly for a  small rescaling of the form $\hbar \rightarrow \hbar(1+\delta)$ where $\delta \ll 1$ \cite{Lake:2018zeg,LakeUkraine2019,Lake:2019nmn,Lake:2020rwc,Lake:2021beh,Lake:2020chb}. 
In this talk, we explore one such model, originally proposed in the series of works \cite{Lake:2018zeg,LakeUkraine2019,Lake:2019nmn,Lake:2020rwc,Lake:2021beh,Lake:2020chb} and explain how it gives rise to an effective dark energy density for the Universe. 
The dark energy density derived from the EGUP is approximately constant over large scales but exhibits spatial oscillations with a wavelength of order 0.1 mm, giving a potentially testable signal that distinguishes it from other models.

The structure of this paper is as follows. 
In Sec. \ref{Sec.2} we outline how rigorously defined versions of the GUP, EUP and EGUP can be obtained by introducing a form of quantum nonlocal geometry as our model of the background space-time. 
In Sec. \ref{Sec.3}, we explain how the field that minimises the product of the generalised position and momentum uncertainties naturally generates an effective dark energy density, matching the observed value $\rho_{\Lambda} = \Lambda c^2/(8\pi G) \simeq 10^{-30}$ ${\rm g \, . \, cm^{-3}}$ \cite{Martin:2012bt} on large scales. 
Conclusions and a brief discussion of prospects for future work are given in Sec. \ref{Sec.4}. 

%%%%%%%%%%%%%%%%%%%%%%%%%%%%%%%%%%%
%%%%%%%%%%%%%%%%%%%%%%%%%%%%%%%%%%%
\section{GURs from Quantum Nonlocal Geometry} \label{Sec.2}

It is straightforward to show that the GUP (\ref{GUP-1}) implies the existence of a minimum length scale, of the order of the Planck length, whereas the EUP (\ref{EUP-1}) implies the existence of a minimum momentum of the order of the de Sitter momentum, 
\begin{eqnarray} \label{MinimumLengthAndMomentumScales}
(\Delta x^i)_{\rm min} \simeq l_{\rm Pl} := \sqrt{\hbar G/c^3} \simeq 10^{-33} \, {\rm cm} \, ,
\nonumber\\
(\Delta p_j)_{\rm min} \simeq m_{\rm dS}c := \hbar\sqrt{\Lambda/3} \simeq 10^{-56} \, {\rm g \, . \, cm \, s^{-1}} \, .
\end{eqnarray}
The latter represents the minimum momentum that a quantum particle can possess when confined within the asymptotic de Sitter horizon, which is comparable to the present day radius of the Universe, 
\begin{eqnarray} \label{deSitterHorizon}
r_{\rm U}(t_0) \simeq l_{\rm dS} := \sqrt{3/\Lambda} \simeq 10^{28} \, {\rm cm} \, . 
\end{eqnarray}

The basic idea of the new model \cite{Lake:2018zeg,LakeUkraine2019,Lake:2019nmn,Lake:2020rwc,Lake:2021beh,Lake:2020chb} is to implement minimum length and momentum scales by introducing a form of quantum nonlocal geometry, in which each point in the classical background space is identified with a complex quantum amplitude density and ‘smeared’ over a region comparable to the Planck volume. 
To this end, each point `$\bold{x}$' in the classical geometry is first associated with the rigged basis vector of a Hilbert space, i.e., a ket $\ket{\bold{x}}$. 
Each precise-valued ket is then smeared by mapping it to a superposition of such quantum `points', with a characteristic width of order $l_{\rm Pl}$. 
This is implemented via the map
\begin{eqnarray} \label{SmearingMap-1}
S : \ket{\bold{x}} \mapsto \ket{\bold{x}} \otimes \ket{g_{\bold{x}}} \, ,  
\end{eqnarray}
where 
\begin{eqnarray} \label{g_x}
\ket{g_{\bold{x}}} = \int g(\bold{x}' - \bold{x}) \ket{\bold{x}'} {\rm d}^{3}{\rm x}' %\, ,
\end{eqnarray}
is any normalised wave function $\braket{g_{\bold{x}}|g_{\bold{x}}} = 1$. 
For simplicity, $|g(\bold{x}' - \bold{x})|^2$ my be thought of as a Planck-width Gaussian, centred on the fixed classical point `$\bold{x}$' \cite{Lake:2018zeg,Lake:2020rwc}. 
Applying the map $S$ (\ref{SmearingMap-1}) to the wave vector of a canonical quantum particle, $\ket{\psi} = \int \psi(\bold{x}) \ket{\bold{x}} {\rm d}^{3}{\rm x}$, gives 
\begin{eqnarray} \label{psi->Psi}
S : \ket{\psi} \mapsto \ket{\Psi} \, , 
\end{eqnarray}
where
\begin{eqnarray} \label{|Psi>_position_space}
\ket{\Psi} = \int\int \psi(\bold{x}) g(\bold{x}' - \bold{x}) \ket{\bold{x},\bold{x}'} {\rm d}^{3}{\rm x}{\rm d}^{3}{\rm x}' \, ,
\end{eqnarray}
and $\ket{\bold{x},\bold{x}'} := \ket{\bold{x}} \otimes \ket{\bold{x}'}$. 
This corresponds to the generalised wave function
\begin{eqnarray} \label{Psi}
\Psi(\bold{x},\bold{x}') = \psi(\bold{x}) g(\bold{x}'-\bold{x}) \, , 
\end{eqnarray}
which incorporates the quantum fluctuations associated with both material particles and the background space on which they propagate. 

The key point is that, due to this smearing, there are two position-type variables in this model. 
It is important to recognise that the first variable, `$\bold{x}$', refers to the quantum state of a geometric ‘point’, and not to the position of a material particle on a fixed classical background, as in canonical quantum theory. 
The second variable, `$\bold{x}'$', represents the observed position of a material particle in the new ‘fuzzy’ space, created by smearing. 
The probability of observing the particle at a given position value, $\bold{x{\, '}}$, given the generalised wave function $\Psi$, is
\begin{eqnarray} \label{GeneralisedXProbability}
\frac{{\rm d}^{3}P(\bold{x}' | \Psi)}{{\rm d}{\rm x}'^{3}} = \int |\Psi(\bold{x},\bold{x}')|^2 {\rm d}^3{\rm x} = (|\psi|^2 * |g|^2)(\bold{x}') \, .
\end{eqnarray}

In other words, new quantum degrees of freedom are introduced to describe Planck scale fluctuations of the background geometry.
These fluctuations depend on the smearing function $g$ (e.g. a Planck-width Gaussian) and not on the canonical quantum wave function $\psi$. 
The variances associated with each source of position uncertainty add linearly, giving a generalised uncertainty relation (GUR) of the form 
\begin{eqnarray} \label{X_uncertainty}
(\Delta_\Psi X^{i})^2 = (\Delta_\psi x'^{i})^2 + (\Delta_gx'^i)^2 \, ,
\end{eqnarray}
where $\hat{\bold{X}} := \bold{x} + (\bold{x}' - \bold{x}) \equiv \bold{x}'$ is the generalised position operator in the position space representation of the wave mechanics picture. Whereas $\psi$ controls the fluctuations in $\bold{x}$, $g$ controls the fluctuations in $\bold{x}' - \bold{x}$, and may be interpreted as the complex quantum amplitude for the coherent transition $\bold{x} \leftrightarrow \bold{x}'$ \cite{Lake:2019nmn,Lake:2020rwc}. 
Fixing the width of $|g|^2$ to be of the order of the Planck length, $\sigma_g \simeq l_{\rm Pl}$, Eq. (\ref{X_uncertainty}) Taylor expands, to first order, to yield the GUP-type uncertainty relation
\begin{eqnarray} \label{smeared_GUP}
\Delta_\Psi X^{i} \gtrsim \frac{\hbar}{2\Delta_{\psi} p'_{j}}\delta^{i}{}_{j}\left[1 + \frac{4G}{c^3}(\Delta_{\psi} p'_{j})^2\right] \, . 
\end{eqnarray}
%In this expression, $(\Delta_{\psi} p'_{j})^2 = (\Delta_{\psi} p'_{j}) \, . \, (\Delta_{\psi} p'^{j})$, but no sum is implied by the repeated index.

The formalism outlined above can be extended to the momentum space representation, giving the generalised momentum space wave function \cite{Lake:2018zeg,Lake:2019nmn,Lake:2020rwc}
\begin{eqnarray} \label{tildePsi}
\tilde{\Psi}(\bold{p},\bold{p}') = \tilde{\psi}_{\hbar}(\bold{p}) \, \tilde{g}_{\beta}(\bold{p}'-\bold{p}) \, .
\end{eqnarray}
The associated probability density is 
\begin{eqnarray} \label{GeneralisedPProbability}
\frac{{\rm d}^{3}P(\bold{p}' | \tilde{\Psi})}{{\rm d}{\rm p}'^{3}} = \int |\tilde{\Psi}(\bold{p},\bold{p}')|^2 {\rm d}^{3}{\rm p} = (|\tilde{\psi}_{\hbar}|^2 * |\tilde{g}_{\beta}|^2)(\bold{p}') \, ,
\end{eqnarray}
giving an uncertainty relation of the form
\begin{eqnarray} \label{P_uncertainty}
(\Delta_\Psi P_{j})^2 = (\Delta_\psi p'_{j})^2 + (\Delta_g p'_j)^2 \, ,
\end{eqnarray}
where $\hat{\bold{P}} := \bold{p} + (\bold{p}' - \bold{p}) \equiv \bold{p}'$ is the generalised momentum operator in the momentum space representation of the wave mechanics picture.

Here, $\tilde{\psi}_{\hbar}(\bold{p})$ denotes the $\hbar$-weighted Fourier transform of the position space wave function, $\psi(\bold{x})$, as in canonical quantum mechanics, 
\begin{eqnarray} \label{CanonicaldeBroglie}
\tilde{\psi}_{\hbar}(\bold{p}) = \left(\frac{1}{\sqrt{2\pi\hbar}}\right)^3 \int \psi(\bold{x}) e^{-\frac{i}{\hbar}\bold{p}.\bold{x}}{\rm d}^{3}{\rm x} \, . 
\end{eqnarray}
Imposing Eq. (\ref{CanonicaldeBroglie}) is equivalent to imposing the canonical de Broglie relation, $\bold{p} = \hbar \bold{k}$. 
This, in turn, is equivalent to imposing two independent physical assumptions. 
The first is the principle of quantum superposition, which requires $\bold{p} \propto \bold{k}$, and the second is the assumption that $\hbar = 1.05 \times 10^{-34}$ Js determines the scale of wave-particle duality. 
The second assumption determines the normalisation constant, $(\sqrt{2\pi\hbar})^{-3}$. 

Therefore, imposing a similar relation on the wave function $g(\bold{x}'-\bold{x})$ is equivalent to assuming that $\hbar$ represents the fundamental quantum of action for quantised geometry, and hence also for gravity in the relativistic limit, as well as for material particles. 
Although virtually all attempts to construct a quantum theory of gravity proposed over the past 90 years tacitly assume this we note that this assumption has, at present, no empirical justification. 
In fact, making such an assumption in our current approach automatically leads to a minimum possible momentum uncertainty of the order of the Planck momentum, which is obviously at odds with empirical observations. 
Alternatively, if we set 
\begin{eqnarray} \label{ModifieddeBroglie}
\tilde{g}_{\beta}(\bold{p}'-\bold{p}) = \left(\frac{1}{\sqrt{2\pi\beta}}\right)^3 \int g(\bold{x}'-\bold{x}) e^{-\frac{i}{\beta}(\bold{p}'-\bold{p}).(\bold{x}'-\bold{x})}{\rm d}^{3}{\rm x}' \, , 
\nonumber
\end{eqnarray}
where
\begin{eqnarray} \label{beta}
\beta \simeq \hbar\sqrt{\frac{\rho_\Lambda}{\rho_{\rm Pl}}} \simeq \hbar \times 10^{-61}% \, , 
\end{eqnarray}
and $\rho_{\rm Pl} := c^5/(\hbar G^2)$ is the Planck density, the width of $|\tilde{g}_{\beta}|^2$ in momentum space is of the order of the de Sitter momentum, $\tilde{\sigma}_{g} \simeq m_{\rm dS}c$. 
This gives the EUP-type uncertainty relation
\begin{eqnarray} \label{smeared_EUP}
\Delta_\Psi P_{j} \gtrsim \frac{\hbar}{2\Delta_{\psi} x'^{i}}\delta^{i}{}_{j}\left[1 + \frac{\hbar\Lambda}{12}(\Delta_{\psi} x'^{i})^2\right] \, ,
\end{eqnarray}
after Taylor expanding Eq. (\ref{P_uncertainty}) to first order. 
Directly combining Eqs. (\ref{X_uncertainty}) and (\ref{P_uncertainty}) together with the HUP (\ref{HUP-2}), which holds independently for the canonical distribution $|\psi|^2$, and again Taylor expanding the resulting expression to first order, yields the EGUP
\begin{eqnarray} \label{smeared-spaceEGUP-2}
\Delta_{\Psi} X^{i} \Delta_{\Psi} P_{j} \gtrsim \frac{\hbar}{2}\delta^{i}{}_{j}\left[1 + \alpha(\Delta_{\Psi} P_{j})^2 + \eta(\Delta_{\Psi} X^{i})^2\right] \, ,
\end{eqnarray}
where $\alpha = 4G/(\hbar c^3)$ and $\eta = \Lambda/6$. 
%\begin{eqnarray} \label{smeared-spaceEGUP-3}
%\alpha = \frac{4G}{\hbar c^3} \, , \quad \eta = \frac{\Lambda}{6} \, .
%\end{eqnarray}

In this way, the quantum wave function of a geometric `point' can be smeared out over a volume of the order of the Planck volume, without generating the kind of energy densities that typically lead to black hole formation \cite{Lake:2020chb}. 
In Sec. \ref{Sec.3}, we will show how this also allows us to evade the so called `worst prediction in physics', namely, the enormous difference between the naive prediction of the vacuum energy and the observed dark energy density \cite{Martin:2012bt}. 
%\cite{Hobson:2006se,Martin:2012bt}

Before concluding this section, however, we highlight an essential aspect of the smeared space model, which distinguishes it from nearly all other GUR models proposed in the existing quantum gravity literature (see \cite{Bishop:2019yft,Bishop:2020cep} for a notable exception). 
It is straightforward to show that the GURs (\ref{X_uncertainty}) and (\ref{P_uncertainty}) can be obtained as the standard deviations of generalised position and momentum operators, $\hat{X}^{i}$ and $\hat{P}_{j}$, satisfying the commutation relations \cite{Lake:2018zeg,Lake:2019nmn,Lake:2020rwc}
\begin{equation} \label{[X,P]}
[\hat{X}^{i},\hat{P}_{j}] = i\hbar(1 + \delta) \, \delta^{i}{}_{j} \, {\bf\hat{\mathbb{I}}} \, ,
\end{equation}
%and 
\begin{equation} \label{}
[\hat{X}^{i},\hat{X}^{j}] = 0 \, , \quad [\hat{P}_{i},\hat{P}_{j}] = 0 \, , 
\end{equation}
where
\begin{eqnarray} \label{delta}
\delta := \hbar/\beta \simeq 10^{-61} \, , 
\end{eqnarray}
which act on the generalised probability distribution $|\Psi|^2 = |\psi|^2|g|^2$, i.e., such that
\begin{eqnarray} \label{}
(\Delta_\Psi X^{i})^2 &=& \braket{\Psi |(\hat{X}^{i})^{2}|\Psi} - \braket{\Psi|\hat{X}^{i}|\Psi}^2 \, , 
\nonumber\\
(\Delta_\Psi P_{j})^2 &=& \braket{\Psi |(\hat{P}_{j})^{2}|\Psi} - \braket{\Psi|\hat{P}_{j}|\Psi}^2 \, . 
\end{eqnarray}

Viewed in this way, the whole formalism of the smeared space model is equivalent to imposing the modified de Broglie relation, 
\begin{eqnarray} \label{ModifieddeBroglie*}
\bold{p}' = \hbar \bold{k} + \beta (\bold{k}' - \bold{k}) \, ,
\end{eqnarray} 
where $\beta$ is the fundamental quantum of action for space-time \cite{Lake:2018zeg,LakeUkraine2019,Lake:2019nmn,Lake:2020rwc,Lake:2021beh,Lake:2020chb}. 
Heuristically, the additional term can be understood as describing the additional momentum ‘kick’ imparted to a material particle, due to quantum fluctuations of the background geometry 
\footnote{Theoretically, a strong case has been made against the existence of multiple quantisation constants for different species of material particles, via a series of no-go theorems (see \cite{Sahoo} and references therein). Technically, what these theorems rule out is the existence of multiple de Broglie relations of the form $\bf{p} = \hbar\bf{k}$ and $\bf{p}' = \hbar'\bf{k}'$, where $\hbar' \neq \hbar$, for different particles species. If one includes gravitons, via the usual linearisation of the Einstein field equations, then these arguments can also be extended to cover the case of the gravitational field \cite{Deser:2022lmi}. However, the standard linearisation procedure $g_{\mu\nu} \simeq \eta_{\mu\nu} + h_{\mu\nu}$, where the Minkowski metric $\eta_{\mu\nu}$ is used to raise and lower the indices of the perturbations $h_{\mu\nu}$, treats the background geometry as fixed and classical. This is equivalent to regarding it as a classical reference frame from which the quantised perturbations (gravitons) can be observed through a series of idealised measurements. In a true quantum gravity scenario we expect the Planck-scale delocalisation of `points' in the background geometry to render such a description untenable and for classical reference frames to be replaced by quantum reference frames (QRF). It is therefore intriguing that in the limit $\beta \rightarrow \hbar$ the formalism presented in  \cite{Lake:2018zeg,Lake:2019nmn,Lake:2020rwc} reduces to the QRF formalism derived by Giacomini, Castro-Ruiz and Brukner in \cite{Giacomini:2017zju}. This suggests that the `smearing' procedure devised therein treats the quantum background geometry as a QRF rather than as a standard quantum field embedded in a sharp classical space-time. Mathematically, this is expressed by the fact that the new quantisation constant $\beta \neq \hbar$ is associated with {\it relative} variables via the modified de Broglie relation $\bf{p}' = \hbar \bf{k} + \beta(\bf{k}'-\bf{k})$ (\ref{ModifieddeBroglie*}). Note that this kind of quantisation condition is {\it not} ruled out by the existing no-go theorems against multiple quantisation constants \cite{Sahoo,Deser:2022lmi}. These issues are discussed in greater detail in \cite{Lake:2020rwc}.}.

Since the resulting commutation relation between the generalised  $\hat{X}^{i}$ and $\hat{P}_{j}$ operators is simply a rescaled version of the standard Heisenberg commutator, with $\hbar \rightarrow \hbar + \beta$, 
this implies, immediately, that all four of the major problems associated with ‘standard’ modified commutator models are avoided. 
In the smeared space formalism, they simply don’t occur. 
This is the most important result of the recent works \cite{Lake:2018zeg,LakeUkraine2019,Lake:2019nmn,Lake:2020rwc,Lake:2021beh,Lake:2020chb}, which demonstrate that GURs can be derived from alternative mathematical structures, corresponding to new physical assumptions, without the need for modified commutation relations of a non-Heisenberg type. 

%%%%%%%%%%%%%%%%%%%%%%%%%%%%%%%%%%%
%\subsection{Section 2.1}

%%%%%%%%%%%%%%%%%%%%%%%%%%%%%%%%%%%
%%%%%%%%%%%%%%%%%%%%%%%%%%%%%%%%%%%
\section{Dark energy from Quantum Nonlocal Geometry} \label{Sec.3}

In this section, we outline the possible consequences of the smeared space model for the observed vacuum energy of the Universe. 
To begin, we note that the EGUP (\ref{smeared_EUP}) is approximate and that, as an intermediate step in its derivation, we obtain the exact expression
\begin{eqnarray} \label{smeared-spaceEGUP-1}
(\Delta_{\Psi} X^{i})^2 (\Delta_{\Psi} P_{j})^2 &\geq& (\hbar/2)^2(\delta^{i}{}_{j})^2 + (\Delta_{\psi}x'^{i})^2(\Delta_{g} p'_{j})^2 
\nonumber\\
&+& (\Delta_{g} x'^{i})^2\frac{ (\hbar/2)^2}{(\Delta_{\psi} x'^{j})^2} + (\Delta_{g}x'^{i})^2(\Delta_{g} p'_{j})^2 \, ,
\nonumber
\end{eqnarray}
plus an analogous relation involving only $(\Delta_{\psi} p'_{j})^2$. 
These are obtained by directly combining Eqs. (\ref{X_uncertainty}) and (\ref{P_uncertainty}), together with the HUP (\ref{HUP-2}), without making any approximations. 
Extremising the product of the generalised uncertainties on the right-hand side of this relation, with respect to $\Delta_\psi x'^{i}$, and its counterpart with respect to $\Delta_\psi p'_{j}$, yields
\begin{equation} \label{EQ_CAN_DX_OPT}
(\Delta_\psi x'^{i})_{\mathrm{opt}} = \sqrt{\frac{\hbar}{2} \frac{\Delta_{g} x'^{i}}{\Delta_{g} p'_{i}}} \, , \quad (\Delta_\psi p'_{j})_{\mathrm{opt}} = \sqrt{\frac{\hbar}{2} \frac{\Delta_{g} p'_{j}}{\Delta_{g} x'^{j}}} \, , 
\end{equation}
giving the inequality
\begin{eqnarray} \label{DXDP_opt}
\Delta_\Psi X^{i} \, \Delta_\Psi P_{j} & \ge & \frac{\hbar \, (1 + \delta)}{2} \, \delta^{i}{}_{j} \, .
\end{eqnarray}
This result also follows directly from the commutator (\ref{[X,P]}), via the Schr{\" o}dinger--Robertson relation. 
Next, we again impose the conditions
\begin{eqnarray} \label{sigma_g}
\sigma_g \simeq l_{\rm Pl} \, , \quad \tilde{\sigma}_g \simeq m_{\rm dS}c \, , 
\end{eqnarray}
where $\sigma_g$ represents the width of $|g|^2$ in position space and $\tilde{\sigma}_g$ denotes the width of $|\tilde{g}_{\beta}|^2$ in momentum space, giving
\begin{eqnarray} \label{}
(\Delta_{\Psi} X)_{\rm opt} \simeq l_{\Lambda} := \sqrt{l_{\rm Pl}l_{\rm dS}} \simeq 0.1 \, {\rm mm} \, , 
\nonumber\\
(\Delta_{\Psi} P)_{\rm opt} \simeq m_{\Lambda}c := \sqrt{m_{\rm Pl}m_{\rm dS}}c \simeq 10^{-3} \, {\rm eV/c} \, .
\end{eqnarray}
Here, we have neglected the directional indices, since these relations hold for the spread of position and momentum in all directions. 
The resulting energy density is
\begin{eqnarray} \label{min_energy_density}
\mathcal{E}_\Psi \simeq \frac{3}{4\pi}\frac{(\Delta_\Psi P)_{\rm opt}\ c}{(\Delta_\Psi X)^3_{\rm opt}} \simeq \rho_{\Lambda}c^2 = \frac{\Lambda c^4}{8\pi G} \, .
\end{eqnarray}

The energy density of any field that optimises the smeared space GURs necessarily generates a contribution to the vacuum energy of the order of the observed dark energy density, on large scales, but should exhibit periodic variations on small scales of order 0.1 mm. 
It is therefore intriguing that tentative evidence for such small-scale oscillations in the Newtonian gravitational force, with approximately this wavelength, has already been observed \cite{Perivolaropoulos:2016ucs,Antoniou:2017mhs}.

In this model vacuum modes seek to optimise the GURs induced by the quantum fluctuations of both material particles and the background geometry on which they propagate yielding the observed vacuum energy 
\footnote{This may be contrasted with the case of most dark energy models, in which only the modes of matter fields contribute to the vacuum energy \cite{Amendola:2015ksp}, and to the case of agegraphic dark energy models, in which fluctuations of the the space-time alone generate the required energy density \cite{Cai:2007us,Wei:2007ty,Kim:2008hz,Pankaj:2022lnk}.}. 
Any attempt to excite higher-order modes leads to increased pair-production of neutral dark energy particles, of mass $m_{\rm \Lambda} \simeq 10^{-3} \, {\rm eV/c^2}$, together with the corresponding expansion of space required to accommodate them, rather than an increase in energy density \cite{Burikham:2015nma,Lake:2017ync,Lake:2017uzd,Lake:2018zeg,Lake:2020rwc,Lake:2020chb,Hashiba:2018hth}. The vacuum energy remains approximately constant over large distances but exhibits granularity on scales comparable to the Compton wavelength of the dark energy field, $l_{\rm \Lambda} \simeq 0.1 \, {\rm mm}$. 

A species of charge-neutral Majorana fermions, with a mass comparable to that of an electron neutrino, may suffice to give rise to this kind of phenomenology, since the positive rest-mass of such particles is exactly balanced by their negative gravitational energy. 
In this way, the Universe can expand {\it ad infinitum}, without violating the conservation of energy \cite{InPreparation}. 
Thus, dark energy is sourced by a more-or-less well understood form of matter, as in standard dynamical dark energy models \cite{Amendola:2015ksp}, but the cosmological constant emerges as an {\it effective} description over large scales due to the induced minimum curvature of order $\sim \Lambda$
\footnote{Here, the term `large' refers to any scale significantly greater than 0.1 mm. The existence of such a minimum curvature could therefore still be verified `locally' on astrophysical scales, i.e., within the Solar System, for example using gravitational lensing \cite{Zhang:2021ygh}.}.

These assumptions correspond to imposing UV and IR cut-offs in the usual estimate of the vacuum energy, $k_{\rm UV} = k_{\Lambda} := 2\pi/l_{\Lambda}$ and $k_{\rm IR} = k_{\rm dS} := 2\pi/l_{\rm dS}$, giving
\begin{eqnarray} \label{rho_vac}
\rho_{\rm vac} \simeq \frac{\hbar}{c} \int^{k_{\rm UV}}_{k_{\rm IR}} \sqrt{k^2 + \left(\frac{m c}{\hbar}\right)^2} {\rm d}^3k  \simeq \rho_{\Lambda}c^2 = \frac{\Lambda c^2}{8\pi G} \, , 
%\simeq \rho_{\rm Pl} := \frac{c^5}{\hbar G^2} \simeq 10^{93} \, {\rm g \, . \, cm^{-3}} \, . 
\end{eqnarray}
with $m \simeq m_{\Lambda}$. 
Taking the usual assumption of a Planck scale cutoff, $k_{\rm UV} = k_{\rm Pl} := 2\pi/l_{\rm Pl}$ \cite{Martin:2012bt}, instead yields $\rho_{\rm vac} \simeq \rho_{\rm Pl} := c^5/(\hbar G^2) \simeq 10^{93}$ ${\rm g \, . \, cm^{-3}}$, which differs by a factor of $\delta^2 \simeq 10^{-122}$ from the observed value. 
In the smeared space model, the difference between this naive estimate and the true dark energy density arises, ultimately, from the difference between the quantisation scales for matter and geometry, $\hbar$ and $\beta$. 
%\textcolor{blue}{This difference may be expressed in terms of a new parameter, i.e. $\Lambda$, but it is important to recognise that $\hbar$ and $\beta$ are the fundamental physical quantities. 
The naive calculation assumes that Planck-length fluctuations are associated with Planck-mass energy densities, but this is not necessarily the case if geometry must be quantised on a different scale to matter, that is, with a difference quantum of action $\beta \ll \hbar$ \cite{Lake:2020chb}.

%%%%%%%%%%%%%%%%%%%%%%%%%%%%%%%%%%%
%%%%%%%%%%%%%%%%%%%%%%%%%%%%%%%%%%%
\section{Conclusions}  \label{Sec.4}

We have shown how the most common and well studied types of generalised uncertainty relation considered in the existing literature, the GUP, EUP and EGUP, can be derived from a rigorous quantum formalism that describes a superposition of fluctuating background geometries \cite{Lake:2018zeg,LakeUkraine2019,Lake:2019nmn,Lake:2020rwc,Lake:2021beh,Lake:2020chb}. 
Mapping the quantum state of each spatial point to a superposition maps a single geometry to a superposition of geometries, as expected in any would be theory of quantum gravity \cite{DeWittMorette:2011zz,Marletto:2017pjr,Lake:2018zeg}. 
In this way, each classical point is `smeared' over a volume comparable to the Planck volume and the resulting {\it quantum} geometry is nonlocal. 

The total uncertainties in these formulae are not heuristic quantities but well defined standard deviations, derived from a generalised probability distribution that incorporates geometric fluctuations as well as the diffusion of the canonical quantum wave function. 
Nonetheless, the resulting generalised commutator is simply a rescaled representation of the canonical Heisenberg commutator, with $h \rightarrow \hbar(1+\delta)$, where $\delta \simeq \sqrt{\hbar^3G\Lambda/c^3} \simeq 10^{-61}$. 
This is a key feature of the model that, crucially, allows it to evade the four main pathologies that plague GUP models based on modified commutation relations of non-Heisenberg type \cite{Hossenfelder:2012jw,Tawfik:2014zca,Lake:2020rwc}. 

Additional implications of the model include the existence of a minimum energy density in nature, i.e. dark energy, which emerges as a logical necessity of the smeared space formalism, without which its momentum space representation is not well defined \cite{Lake:2018zeg}, and the idea that space-time must quantised on a different scale to matter, that is, with a different quantisation constant, $\beta \simeq \hbar \times 10^{-61}$. 
In this formalism the fundamental quantum of action for the geometric degrees of freedom is many orders of magnitude smaller than Planck’s constant. 
Therefore, although space-time is still fundamentally quantum in nature, it behaves, in some respects, much more classically than matter. 
This may have important consequences for the physics of quantised gravitational waves and for future theories that attempt to unify matter and geometry at the epoch of the Big Bang. 
These topics will be studied, in detail, in future works. 

%%%%%%%%%%%%%%%%%%%%%%%%%%%%%%%%%%%
%%%%%%%%%%%%%%%%%%%%%%%%%%%%%%%%%%%
\section{Acknowledgements}

This work was supported by the Guangdong Province Natural Science Foundation, grant no. 008120251030. 
I am grateful to the Frankfurt Institute for Advanced Studies for gracious hospitality during the preparation of the manuscript.

%%%%%%%%%%%%%%%%%%%%%%%%%%%%%%%%%%%
%%%%%%%%%%%%%%%%%%%%%%%%%%%%%%%%%%%

\end{document}